\documentclass[aps,preprint]{revtex4}%
\usepackage{amsfonts}
\usepackage{amsmath}
\usepackage{amssymb}
\usepackage{graphicx}%
\begin{document}
\title[Non-extensive RMT]{Non-extensive RMT Approach to Mixed Regular-Chaotic Dynamics}
\author{A. Y. Abul-Magd}
\affiliation{Department of Mathematics, Faculty of Science, Zagazig University, Zagazig, Egypt}

\begin{abstract}
We apply Tsallis's $q$-indexed entropy to formulate a non-extensive random
matrix theory (RMT), which may be suitable for systems with mixed
regular-chaotic dynamics. The joint distribution of the matrix elements is
given by folding the corresponding quantity in the conventional random matrix
theory by a distribution of the inverse matrix-element variance. It keeps the
basis invariance of the standard theory but violates the independence of the
matrix elements. We consider the sub-extensive regime of $q$ more than unity
in which the transition from the Wigner to the Poisson statistics is expected
to start. We calculate the level density for different values of the entropic
index. Our results are consistent with an analogous calculation by Tsallis and
collaborators. We calculate the spacing distribution for mixed systems with
and without time-reversal symmetry. Comparing the result of calculation to a
numerical experiment shows that the proposed non-extensive model provides a
satisfactory description for the initial stage of the transition from chaos
towards the Poisson statistics.

\end{abstract}
\volumeyear{year}
\volumenumber{number}
\issuenumber{number}
\eid{identifier}
\date{\today}
\startpage{1}
\endpage{2}
\maketitle

\section{INTRODUCTION}

The past decade has witnessed a considerable interest devoted to
non-conventional statistical mechanics. Much work in this direction followed
the line initiated by Tsallis' seminal paper \cite{Ts1}. The standard
statistical mechanics is based on the Shannon entropy measure $S=-\Sigma
_{i}p_{i}\ln p_{i\text{ }}$(we use Boltzmann's constant $k_{B}=1$), where
$\{p_{i}\}$ denotes the probabilities of the microscopic configurations. This
entropy is extensive. For a composite system $A+B$, constituted of two
independent subsystems $A$ and $B$ such that the probability
$p(A+B)=p(A)p(B),$ the entropy of the total $S(A+B)=S(A)+S(B)$. Tsallis
proposed a non-extensive generalization: $S_{q}=\left(  1-\Sigma_{i}p_{i}%
^{q}\right)  /(q-1)$. The entropic index $q$ characterizes the degree of
extensivity of the system. The entropy of the composite system $A+B$, the
Tsallis' measure verifies
\begin{equation}
S_{q}(A+B)=S_{q}(A)+S_{q}(B)+(1-q)S_{q}(A)S_{q}(B),
\end{equation}
from which the denunciation non-extensive comes. Therefore, $S_{q}%
(A+B)<S_{q}(A)+S_{q}(B)$ if $q>1$. This case is called sub-extensive. If
$q<1$, the system is in the super-extensive regime. The standard statistical
mechanics recovered for $q$ = 1. Applications of the Tsallis formalism covered
a wide class of phenomena; for a review please see, e.g. \cite{Ts2}. However,
the relation between the parameter $q$ and the underlying microscopic dynamics
is not fully understood yet. The value of $q$ has been obtained from studies
of dynamics in cases of low-dimensional dissipative maps \cite{TsM1,TsM2}, and
in some toy models of self-organized criticality \cite{TsSO}. Explicit
expressions for $q$ in terms of physical quantities exist in few cases, e.g.
in turbulence problems \cite{beck} and physics of the solar plasma \cite{kan}.
Aringazian and Mazhitov \cite{ari} obtained a Tsallis distribution function
for a smaller subsystem weakly interacting with the remaining
\textquotedblright quasi-thermostat\textquotedblright\ composed of a larger
number $M$ of particles, with an entropic index $q-1\backsim1/M$.

A number of recent publications considered the possibility of a non-extensive
generalization to the random matrix theory (RMT) \cite{mehta}. This is the
statistical theory of random matrices $H$ whose entries fluctuate as
independent Gaussian random numbers. The matrix-element distribution has been
obtained by extremizing Shannon's entropy subject to the constraint of
normalization and existence of the expectation value of Tr$\left(  H^{\dagger
}H\right)  $ \cite{balian}. What has become known as the
Bohigas-Giannoni-Schmidt conjecture is that the quantum spectra of classically
chaotic systems are correlated according to RMT, whereas the spectral
correlations of classically regular systems are close to Poissonian statistics
\cite{bohigas}. Several attempts have been done to extend the applicability of
RMT to include quantum systems with mixed regular-chaotic classical dynamics;
for a review please see \cite{guhr}. For example, the principle of maximum
entropy was used in for this purpose by introducing additional constraints
concerning the off-diagonal elements \cite{hussein}. Non-extensive
generalizations of RMT, on the other hand, extremize Tsallis' non-extensive
entropy, rather than Shannon's. The first attempt in this direction is
probably due to Evans and Michael \cite{evans}. Toscano et al. \cite{toscano}
constructed non-Gaussian ensemble by minimizing Tsallis' entropy and obtained
expressions for the level densities and spacing distributions. Bertuola et al.
\cite{bertuola1} have shown that Tsallis' statistics interpolates between RMT
and an ensemble of L\'{e}vy matrices \cite{cizeau} that have wide range of
applications. They illustrated the spectral fluctuations in the sub-extensive
regime by considering the gap function $E(s)$ that gives the probability of
finding an eigenvalue-free segment of length $s$. Analytical expressions for
the level-spacing distributions of mixed systems belonging to the three
symmetry universality classes are obtained in \cite{abul}. A slightly
different application of non-extensive statistical mechanics to RMT is due to
Nobre et al. \cite{nobre}.

In this work, we use the integral representation of the gamma function to
express the characteristics of the proposed non-extensive RMT in terms of
integrals involving the characteristics of the conventional theory. We show
that non-extensive statistics provides a principled way to accommodate systems
with mixed regular-chaotic dynamics.

\section{NON-EXTENSIVE GENERALIZATION OF RMT}

RMT replaces the Hamiltonian of the system by an ensemble of Hamiltonians
whose matrix elements are independent random numbers. Dyson \cite{dyson}
showed that there are three generic ensembles of random matrices, defined in
terms of the symmetry properties of the Hamiltonian. Time-reversal-invariant
quantum system are represented by a Gaussian orthogonal ensemble (GOE) of
random matrices when the system has rotational symmetry and by a Gaussian
symplectic ensemble (GSE) otherwise. Chaotic systems without time reversal
invariance are represented by the Gaussian unitary ensemble (GUE). The
dimension $\beta$ of the underlying parameter space is used to label these
three ensembles: for GOE, GUE and GSE, $\beta$ takes the values 1, 2 and 4,
respectively. Balian \cite{balian} derived the weight functions $P_{\beta}(H)$
for the three Gaussian ensembles from the maximum entropy principle
postulating the existence of a second moment of the Hamiltonian. He applied
the conventional Shannon definition for the entropy to ensembles of random
matrices as $S=-\int dHP_{\beta}(H)\ln P_{\beta}(H)$ and maximized it under
the constraints of normalization of $P_{\beta}(H)$\ and fixed mean value of
Tr$\left(  H^{\dagger}H\right)  .$ He obtained $P_{\beta}(H)\varpropto
\exp\left[  -\eta Tr\left(  H^{\dagger}H\right)  \right]  $\ which is a
Gaussian distribution with inverse variance $1/2\eta$. In this section, we
apply the maximum entropy principle, with Tsallis' entropy, to random-matrix
ensembles belonging to the three canonical symmetry universalities. The
Tsallis entropy is defined for the joint matrix-element probability density
$P_{\beta}(q,H)$\ by
\begin{equation}
S_{q}\left[  P_{\beta}(q,H)\right]  =\left.  \left(  1-\int dH\left[
P_{\beta}(q,H)\right]  ^{q}\right)  \right/  (q-1).
\end{equation}
We shall refer to the corresponding ensembles as the Tsallis orthogonal
ensemble (TsOE), the Tsallis Unitary ensemble (TsUE), and the Tsallis
symplectic ensemble (TsSE). For $q\rightarrow1$, $S_{q}\ $tends to Shannon's
entropy, which yields the canonical Gaussian orthogonal, unitary or symplectic
ensembles\ (GOE, GUE, GSE) \cite{mehta,balian}.

There are more than one formulation of non-extensive statistics which mainly
differ in the definition of the averaging. Some of them are discussed in
\cite{wang}. We apply the most recent formulation \cite{Ts3}. The probability
distribution $P_{\beta}(q,H)$\ is obtained by maximizing the entropy under two
conditions,
\begin{align}
\int dHP_{\beta}(q,H)  &  =1,\\
\frac{\int dH\left[  P_{\beta}(q,H)\right]  ^{q}\text{Tr}\left(  H^{\dagger
}H\right)  }{\int dH\left[  P_{\beta}(q,H)\right]  ^{q}}  &  =\sigma_{\beta
}^{2}%
\end{align}
where $\sigma_{\beta}$ is a constant. The optimization of $S_{q}$ with these
constraints yields a power-law type for $P_{\beta}(q,H)$
\begin{equation}
P_{\beta}(q,H)=\widetilde{Z}_{q}^{-1}\left[  1+(q-1)\widetilde{\eta}%
_{q}\left\{  \text{Tr}\left(  H^{\dagger}H\right)  -\sigma_{\beta}%
^{2}\right\}  \right]  ^{-\frac{1}{q-1}},
\end{equation}
where $\widetilde{\eta}_{q}>0$ is related to the Lagrange multiplier $\eta$
associated with the constraint in (4) by
\begin{equation}
\widetilde{\eta}_{q}=\eta/\int dH\left[  P_{\beta}(q,H)\right]  ^{q},
\end{equation}
and\
\begin{equation}
\widetilde{Z}_{q}=\int dH\left[  1+(q-1)\widetilde{\eta}_{q}\left\{
\text{Tr}\left(  H^{\dagger}H\right)  -\sigma_{\beta}^{2}\right\}  \right]
^{-\frac{1}{q-1}}.
\end{equation}
It turns out that the distribution (5) can be written hiding the presence of
$\sigma_{\beta}^{2}$ in a more convenient form
\begin{equation}
P_{\beta}(q,H)=Z_{q}^{-1}\left[  1+(q-1)\eta_{q}\text{Tr}\left(  H^{\dagger
}H\right)  \right]  ^{-\frac{1}{q-1}},
\end{equation}
where
\begin{equation}
\eta_{q}=\frac{\eta}{\int dH\left[  P_{\beta}(q,H)\right]  ^{q}+(1-q)\eta
\sigma_{\beta}^{2}},
\end{equation}
and
\begin{equation}
Z_{q}=\int dH\left[  1+(q-1)\eta_{q}\text{Tr}\left(  H^{\dagger}H\right)
\right]  ^{-\frac{1}{q-1}}.
\end{equation}
The non-extensive distribution (8) is reduced to the statistical weight of the
Gaussian ensemble when $q=1$.

It is important to note that the non-extensive distribution $P_{\beta}%
(q,\eta_{q},H)$ is isotropic in the Hilbert space because the dependence on
the matrix elements of $H$ enters through Tr$\left(  H^{\dagger}H\right)  $.
In this way, Tsallis' statistics offers a random-matrix model for mixed
systems, which is invariant under change of basis unlike most of the models in
the literature. However, the distribution does not factorize into a product of
distributions corresponding to the individual matrix elements if $q\neq1$.
Physically, this implies that the starting hypothesis of the standard RMT that
the matrix elements are independent random variable does not hold in the
non-extensive context described by Eq. (2).

The formalism developed in this section was applied in Ref. \cite{abul} to
ensembles of $2\times2$ matrices. The calculation of the spacing distribution
showed different behavior depending on whether $q$ is above or below 1. It is
found that the sub-extensive regime of $q>1$ corresponds to the evolution of a
mixed system towards a state of order described by the Poisson statistics. On
the other hand, the spectrum in the super-extensive regime develops towards
the picked-fence type, such as the one obtained by Berry and Tabor
\cite{berry} for the two-dimensional harmonic oscillator with non-commensurate frequencies.

\subsection{Sub-extensive regime}

In this paper, we shall consider only the sub-extensive regime, where $q>1$.
We note that Tr$\left(  H^{\dagger}H\right)  =\sum_{i=1}^{N}\left(
H_{ii}^{(0)}\right)  ^{2}+2\sum_{\gamma=0}^{\beta-1}\sum_{i>j}\left(
H_{ij}^{(\gamma)}\right)  ^{2}$, where all the four matrices $H^{(\gamma)}$
with $\gamma=0,1,2,3$ are real and where $H^{(0)}$ is symmetric while
$H^{(\gamma)}$ with $\gamma=1,2,3$ are antisymmetric. We introduce the new
coordinates $\mathbf{y}=\left\{  y_{1},\cdots,y_{d}\right\}  $, where
$d=N+\beta N(N-1)/2$, and $y_{i}^{2}$ stand for the square of the diagonal
elements or twice the square of the non-diagonal elements, respectively, and
express the integrals in hyperspherical coordinates. The normalization
condition (3) yields%
\begin{multline}
Z_{q}(\eta_{q})=2^{-\beta N(N-1)/2}\Omega_{d}\int_{0}^{\infty}y^{d-1}dy\left[
1+(q-1)\eta_{q}y^{2}\right]  ^{-\frac{1}{q-1}}\\
=2^{-\beta N(N-1)/2}\Omega_{d}\frac{\Gamma\left(  \frac{1}{q-1}-\frac{d}%
{2}\right)  }{2\left[  \left(  q-1\right)  \eta_{q}\right]  ^{d/2}\Gamma
(\frac{d}{2})\Gamma\left(  \frac{1}{q-1}\right)  }%
\end{multline}
provided that $q<1+2/d$\ , otherwise the integral diverges. Here $\Omega
_{d}=2\pi^{d/2}/\Gamma(d/2)$ is the area of a unit $d$-dimensional hypersphere
and $\Gamma(z)$ is Euler's gamma function. Condition (4), which now reads
\begin{equation}
\sigma_{\beta}^{2}=\left.  \int_{0}^{\infty}y^{d+1}dy\left[  1+(q-1)\eta
_{q}y^{2}\right]  ^{-\frac{q}{q-1}}\right/  \int_{0}^{\infty}y^{d-1}dy\left[
1+(q-1)\eta_{q}y^{2}\right]  ^{-\frac{q}{q-1}}.
\end{equation}
The latter yield the following relationship between $\eta_{q}$\ and
$\sigma_{\beta}^{2}$:
\begin{equation}
\eta_{q}=\frac{d}{\sigma_{\beta}^{2}\left(  2+d-dq\right)  }.
\end{equation}
For a Gaussian ensemble, $\sigma_{\beta}^{2}=2dv^{2}$, where $v^{2}$\ is the
variance of each of the non-diagonal matrix elements (or each of their
components), so that $\eta_{1}=1/(4v^{2})$. Condition (4) thus imposes the
following upper limit on $q$%
\begin{equation}
q<1+\frac{2}{d},
\end{equation}
beyond which the non-extensive formalism is not applicable for random matrix
ensembles. This condition prevents the evolution of a chaotic system towards a
state of order from reaching its terminal stage of the Poisson fluctuation
statistics, as we may see in \cite{abul} for the case of $N=2$, and later in
this paper for the general case. The upper limit in (14) is essentially the
extensive limit ($q\rightarrow1+0$) since RMT is meant essentially for large
matrices ($d\rightarrow\infty$). For example, the sub-extensive regime for a
GOE of $20\times20$ matrices is associated with values of $q$ in the narrow
range of $1<q<1.1$. In spite of this, a minor non-extensivity produces a
considerable effect on the spectral statistics of a large system as
demonstrated below. This is attributed to the existence of the non-trivial
"thermodynamic limit" $N(q-1)=$ constant, as pointed out by Botet et al.
\cite{botet}.

\subsection{Integral representation}

In the sub-extensive regime, the non-extensive RMT can arise from the
extensive one by allowing the variances of the matrix elements to fluctuate
using a transformation suggested by Wilk and W\l odarczyk \cite{wilk} and Beck
\cite{beck1}. From Euler's representation of the Gamma function
\cite{abramowitz}, $\Gamma(x)=\int_{0}^{\infty}t^{x-1}e^{-t}dt$, one can
easily derive the following expression
\begin{equation}
\left[  1+(q-1)\eta_{q}\text{Tr}\left(  H^{\dagger}H\right)  \right]
^{-\frac{1}{q-1}}=\frac{1}{\Gamma\left(  \frac{1}{q-1}\right)  }\int
_{0}^{\infty}t^{\frac{1}{q-1}-1}e^{-t\left[  1+(q-1)\eta_{q}\text{Tr}\left(
H^{\dagger}H\right)  \right]  }dt,
\end{equation}
which is possible if $q>1$. We now change the integration variable into
$\eta=(q-1)\eta_{q}t$. The joint distribution of matrix elements of a Tsallis
random-matrix ensemble can then be expressed as
\begin{equation}
P_{\beta}(q,\eta_{q},H)=\int_{0}^{\infty}f_{n}(\eta)\frac{Z_{1}\left(
\eta\right)  }{Z_{q}\left(  \eta_{q}\right)  }P_{\beta}(\eta,H)d\eta,
\end{equation}
where
\begin{equation}
P_{\beta}(\eta,H)=Z_{1}^{-1}e^{-\eta\text{Tr}\left(  H^{\dagger}H\right)  },
\end{equation}
with
\begin{equation}
Z_{1}\left(  \eta\right)  =\int dHe^{-\eta\text{Tr}\left(  H^{\dagger
}H\right)  }=\frac{2^{-\beta N(N-1)/2}\Omega_{d}}{2\eta^{d/2}\Gamma\left(
d/2\right)  }%
\end{equation}
is the distribution function for a Gaussian random-matrix ensemble with
fluctuating matrix-element inverse variance $1/2\eta$, and $f_{n}(\eta)$\ is
the probability density of the $\chi^{2}$-distribution (the distribution of
sum of squares of $n$ normal variables with zero mean and unit variance),
\begin{equation}
f_{n}(\eta)=\frac{1}{\Gamma\left(  n/2\right)  }\left(  \frac{n}{2\eta_{q}%
}\right)  ^{n/2}\eta^{n/2-1}\exp\left(  -\frac{n\eta}{2\eta_{q}}\right)  ,
\end{equation}
with order $n=2/(q-1)$ and mean value $\eta_{q}=nd\left/  [2\sigma_{\beta}%
^{2}(n-d.)]\right.  =n\left/  [4v^{2}(n-d.)]\right.  $. Therefore, the
generalized distribution function $P_{\beta}(q,\eta_{q},H)$\ of non-extensive
statistics is expressed in terms of the distribution function $P_{\beta}%
(\eta,H)$ of the corresponding Gaussian random-matrix ensemble by averaging
over $\eta$, provided that $\eta$ has a $\chi^{2}$ distribution.

As mentioned above, that the non-extensive Hamiltonian matrix-element
distribution in Eqs. (5) and (16) is invariant under change of basis. The mean
value of each matrix element $\left\langle H_{ij}^{(\gamma)}\right\rangle =0$.
On the other hand, the mean value of the square of a matrix element%
\begin{equation}
\left\langle \left(  H_{ij}^{(\gamma)}\right)  ^{2}\right\rangle
=\frac{1+\delta_{ij}}{4}\frac{n}{\eta_{q}(n-d-2)}=\left(  1+\delta
_{ij}\right)  v^{2}\frac{n-d}{n-d-2},
\end{equation}
which is equal to the corresponding quantity in the standard RMT when
$n\rightarrow\infty$, as expected. The distribution does not factorize into a
product of distributions of individual matrix element, or matrix-element
components, as in the standard RMT. The relative dispersion of the squares of
the matrix elements%
\begin{equation}
\frac{\left\langle \left(  H_{ij}^{(\gamma)}\right)  ^{2}\left(  H_{i^{\prime
}j^{\prime}}^{(\gamma^{\prime})}\right)  ^{2}\right\rangle -\left\langle
\left(  H_{ij}^{(\gamma)}\right)  ^{2}\right\rangle \left\langle \left(
H_{i^{\prime}j^{\prime}}^{(\gamma^{\prime})}\right)  ^{2}\right\rangle
}{\left\langle \left(  H_{ij}^{(\gamma)}\right)  ^{2}\right\rangle
\left\langle \left(  H_{i^{\prime}j^{\prime}}^{(\gamma^{\prime})}\right)
^{2}\right\rangle }=.\frac{2}{n-d-4}%
\end{equation}
vanishes only in the extensive limit of $n\rightarrow\infty$. We note that,
for a given $n$ and fixed $v$, the degree of correlation of matrix element
measured by the covariance of their squares increases with increasing the
dimension of the ensemble. This agrees with the result of non-extensive
thermostatistics for the partition function of a system with $N$ subsystems,
which strongly suggest that the factorization approximation fails when $N$ is
large \cite{lenzi}.

\subsection{Eigenvalue distribution}

We now calculate the joint probability density for the eigenvalues of the
Hamiltonian $H$. With $H=U^{-1}XU$, where $U$\ is the global unitary group, we
introduce the elements of the diagonal matrix of eigenvalues $X=$
diag$(x_{1},\cdots,x_{N})$ of the eigenvalues and the independent elements of
$U$ as new variables. Then the volume element (4) has the form
\begin{equation}
dH=\left\vert \Delta_{N}\left(  X\right)  \right\vert ^{\beta}dXd\mu(U),
\end{equation}
where $\Delta_{N}\left(  X\right)  =\prod_{n>m}(x_{n}-x_{m})$ is the
Vandermonde determinant and $d\mu(U)$ the invariant Haar measure of the
unitary group \cite{mehta,guhr}. The probability density $P_{\beta}(H)$ is
taken to be invariant under arbitrary rotations in the matrix space,
$P_{\beta}(\eta,H)=P_{\beta}(\eta,U^{-1}HU)$. \ Integrating over $U$ yields
the joint probability density of eigenvalues in the form
\begin{equation}
P_{\beta}^{(q)}(\eta_{q},x_{1},\cdots,x_{N})=\frac{\Gamma\left(  \frac{n}%
{2}\right)  }{\left(  n/2\right)  ^{d/2}\Gamma\left(  \frac{n-d}{2}\right)
}\int_{0}^{\infty}f_{n}(\eta)\left(  \frac{\eta_{q}}{\eta}\right)
^{d/2}P_{\beta}^{(1)}(\eta,x_{1},\cdots,x_{N})d\eta,
\end{equation}
where $P_{\beta}^{(1)}(\eta,x_{1},\cdots,x_{N})$ is the eigenvalue
distribution of the corresponding Gaussian ensemble, which is given by
\begin{equation}
P_{\beta}^{(1)}(\eta,x_{1},\cdots,x_{N})=C_{\beta}\left\vert \Delta_{N}\left(
X\right)  \right\vert ^{\beta}\exp\left[  -\eta\sum_{i=1}^{N}x_{i}^{2}\right]
,
\end{equation}
where $C_{\beta}$ is a normalization constant. Similar relations can be
obtained for the statistics that can be obtained from $P_{\beta}^{(q)}%
(\eta_{q},x_{1},\cdots,x_{N})$\ by integration.

The $k$-point correlation function \cite{mehta,guhr} measures the probability
density of finding a level near each of the positions $x_{1},\cdots,x_{k}$,
the remaining levels not being observed. It is obtained by integrating the
eigenvalue joint probability density (16) over $N-k$ arguments
\begin{equation}
R_{\beta,k}^{(q)}(\eta_{q},x_{1},\cdots,x_{k})=\int_{-\infty}^{\infty}%
dx_{k+1}\cdots\int_{-\infty}^{\infty}dx_{N}P_{\beta}^{(q)}(\eta,x_{1}%
,\cdots,x_{N}).
\end{equation}
Therefore, the non-extensive generalization of the k-point function of a
Gaussian ensemble $R_{\beta,k}^{(1)}(\eta,x_{1},\cdots,x_{k})$\ is given by
\begin{equation}
R_{\beta,k}^{(q)}(\eta_{q},x_{1},\cdots,x_{k})=\frac{1}{\Gamma\left(  \frac
{m}{2}\right)  }\int_{0}^{\infty}\left(  \frac{n\eta}{2\eta_{q}}\right)
^{\frac{m}{2}}e^{-n\eta/2\eta_{q}}R_{\beta,k}^{(1)}(\eta,x_{1},\cdots
,x_{k})\frac{d\eta}{\eta},
\end{equation}
where%
\[
m=n-d
\]
Noting that $\frac{n}{2\eta_{q}}=2v^{2}m$, we can easily see that the main
parameter is $m$, which is subject to the restriction%
\begin{equation}
0<m<\infty.
\end{equation}
The lower limit follow from the normalization condition as well as the
constraint of finite average matrix norm (4). The upper limit corresponds the
the standard RMT.

\section{LEVEL DENSITY}

The main goal of RMT is to describe the fluctuations of the energy spectra.
Before the study of the fluctuations can be attempted, one must make a
separation between the local level fluctuation from the overall energy
dependence of the level separation. The level density of the standard random
matrix ensembles is not directly related to the physical level density of the
investigated systems. Nevertheless, it is essential to the proper unfolding of
the spectral fluctuation measures. For the $N$-dimensional GOE, the level
density normalized to 1 is given by Wigner's semi-circle law
\begin{equation}
\rho_{1}(\infty,\varepsilon)=\frac{2}{\pi}\sqrt{\eta/N}\sqrt{1-\eta
\varepsilon^{2}/N}.
\end{equation}
Here $\varepsilon=x/v$\ is the energy expressed in units of standard deviation
of the majority of matrix elements. In the following, we derive a
corresponding formula for the non-extensive generalization of GOE, to which we
shall refer as the Tsallis orthogonal ensemble (TsOE).

The level density is obtained by integrating the joint probability density of
eigenvalues over all variables except one, so that the level density of TsOE
$\rho_{q}(m,x)=R_{1,1}^{(q)}(\eta_{q},x)$. Therefore, using Eq. (28) for
$R_{1,1}^{(1)}$ into Eq. (26), we obtain
\begin{equation}
\rho_{q}(m,\varepsilon)=\frac{\left(  2mN\right)  ^{m/2}\Gamma\left(
\frac{m+1}{2}\right)  }{\sqrt{\pi}\Gamma\left(  \frac{m}{2}\right)
\Gamma\left(  \frac{m}{2}+2\right)  }\left\vert \varepsilon\right\vert
^{-m-1}{}_{1}F_{1}\left(  \frac{m+1}{2},\frac{m}{2}+2,-\frac{2mN}%
{\varepsilon^{2}}\right)  \text{,}%
\end{equation}
where $_{1}F_{1}\left(  a,b,z\right)  $ is Kummer's confluent hypergeometric
function \cite{abramowitz}. At $\varepsilon=0$, Eq. (27) yields%
\begin{equation}
\rho_{q}(m,0)=\frac{1}{\pi}\sqrt{\frac{2}{mN}}\frac{\Gamma\left(  \frac
{m+1}{2}\right)  }{\Gamma\left(  \frac{m}{2}\right)  }.
\end{equation}
Using the asymptotic properties of $\Gamma(z)$ at large $z$ \cite{abramowitz},
we can show that the height of $\rho_{q}(m,\varepsilon)$ at the origin is
lower than the GOE level density, $\rho_{1}(\infty,0)=1/\left(  \pi\sqrt
{N}\right)  $, since the ratio of the gamma functions in the right-hand side
of Eq. (30) is approximately equal to $\sqrt{m/2}[1-1/(4m)]$ for $m\gg1$. At
small $m$, where the relation $\Gamma(1+z)=z\Gamma(z)$ tells us that
$\Gamma\left(  m/2\right)  \approx2/m$, the dependence of $\rho_{q}(m,0)$ on
$m$ is mainly given by the factor $\sqrt{m}$. On the other hand, the
asymptotic behavior of $\rho_{q}(m,\varepsilon)$ at large $\left\vert
\varepsilon\right\vert $ is given by%
\begin{equation}
\rho_{q}(m,\varepsilon)\sim\left\vert \varepsilon\right\vert ^{-m-1}.
\end{equation}
Decreasing $m$ from very large values lowers the magnitude of $\rho
_{q}(m,\varepsilon)$ at the origin below the semi-circle form of the Gaussian
ensemble and raises its values at the periphery. This effect is clearly
demonstrated in the left panel Fig. 1. The behavior shown in this figure is
similar to that of the results of calculations by Toscano et al.
\cite{toscano}.

In order to perform the statistical analysis of level fluctuations of the
energy levels, one must take into account that the level density and hence the
level spacing are strongly dependent on the intrinsic energy. For this
purpose, the investigated spectra are transformed into the so-called
'unfolded' spectra \cite{bohigas1} for which the local mean spacing is 1. On
the other hand, calculations using RMT are performed for levels near the
origin, where the level density is nearly equal to a constant proportional to
$\sqrt{\eta}$. The energy scale is so far defined by the standard deviation of
matrix elements, as, e.g. in Eq. (20). It is more suitable to express the
quantities having the dimension of energy in terms of the mean level spacing
rather than standard deviation of matrix elements. For this purpose we replace
the ratio $n\eta/2\eta_{q}$ \ in Eq. (19) by $\eta/\eta_{0}$ and define
$\eta_{0}$\ by the requirement that the mean level spacing\ is 1. We then
obtain for the non-extensive generalization any statistic $R_{\beta
}^{\text{GE}}$ of a Gaussian ensemble
\begin{equation}
R_{\beta}^{(m)}=\frac{1}{\Gamma\left(  \frac{m}{2}\right)  }\int_{0}^{\infty
}\left(  \frac{\eta}{\eta_{0}}\right)  ^{\frac{m}{2}}e^{-\frac{\eta}{\eta_{0}%
}}R_{\beta}^{\text{GE}}(\eta)\frac{d\eta}{\eta},
\end{equation}
where $\eta$ is now understood as the square of the level density.

\section{NEAREST-NEIGHBOR-SPACING DISTRIBUTION}

The nearest-neighbor-spacing distribution (NNSD) is frequently used for the
analysis of experimental spectra. Unfortunately, RMT does not provide a closed
form expression for NNSD. A very good approximation for this distribution is
given by the so-called Wigner surmise \cite{mehta}, which is the exact spacing
distribution for Gaussian ensembles of $2\times2$ matrices. In this section we
shall assume this approximation. We substitute the Wigner surmise for GOE and
GUE into Eq. (32) and obtain expressions for NNSD of the corresponding Tallis ensembles.

\subsection{Systems invariant under time reversal}

Chaotic systems, whose Hamiltonians are invariant under time reversal, are
modeled in RMT by GOE. For these ensembles, the Wigner surmise is
\begin{equation}
P^{\text{GOE}}(\eta,s)=\eta se^{-\frac{1}{2}\eta s^{2}},
\end{equation}
where $\eta$ is obtained by requiring that $P^{\text{GOE}}$ has a mean spacing
equal 1. Substituting (26) into (25), we obtain the following expression for
the non-extensive (Tsallis) orthogonal ensemble
\begin{equation}
P^{\text{TsOE}}(m,s)=\frac{\frac{1}{2}m\eta_{1}s}{\left(  1+\frac{1}{2}%
\eta_{1}s^{2}\right)  ^{1+m/2}},
\end{equation}
where
\begin{equation}
\eta_{1}=\frac{\pi}{2}\left[  \frac{\Gamma\left(  \frac{1}{2}\left(
m-1\right)  \right)  }{\Gamma\left(  \frac{1}{2}m\right)  }\right]  ^{2}%
\end{equation}
is obtained by requiring that $\int_{0}^{\infty}sP^{\text{TsOE}}%
(m,s)ds=1$.\newline

In the limit of $m\rightarrow\infty$, $\eta_{0}\approx\pi/m$ and the
non-extensive NNSD approaches the extensive one, which is peaked at
$s=\sqrt{2/\pi}\approx0.80$. At the other limit where $m=1$, beyond which the
mean the spacing distribution diverges. If one also requires that the second
moment is finite, one increases the lower limit into $m=2$, for which one
obtains $\eta_{0}=\pi^{2}/2$\ and the NNSD becomes
\begin{equation}
P^{\text{TsOE}}(2,s)=\frac{\pi^{2}s}{\left(  1+\pi^{2}s^{2}\right)  ^{2}},
\end{equation}
which vanishes at the origin, has a peak at $s=2\sqrt{2}/\left(  \pi\sqrt
{3}\right)  \approx0.52$, and decays asymptotically as $s^{-3}$. We therefore
see that increasing non-extensivity modifies the NNSD from a Wigner form
towards a Poisson distribution $e^{-s}$\ but never reaches it. Equation (34)
agrees with the corresponding result obtained for $2\times2$ random matrix
ensemble directly by integrating the joint eigenvalue distribution
\cite{abul}. This behavior is explicitly demonstrated in that paper. We thus
expect the non-extensive RMT to describe the transition out of chaos, at least
in its initial stage. We note that our result for the NNSD agrees with the
corresponding distributions obtained in Ref. \cite{toscano} for the case of
orthogonal universality.

Figure 2 compares the NNSD in Eq. (34) with the corresponding results of a
numerical experiment \cite{zyc}. This experiment imitates a one-parameter
(denoted by $\delta$) transition between an ensemble of diagonal matrices with
independently and uniformly distributed elements and a circular orthogonal or
unitary ensemble. The figure shows that the TsOE distributions are generally
in agreement with the numerical-experimental distribution although the quality
of agreement gradually deteriorates, as expected, as the departure from chaos increases.

\subsection{Systems without time reversal symmetry}

RMT models systems whose Hamiltonians violates time reversibility. by GUE. The
corresponding Wigner surmise is
\begin{equation}
P^{\text{GUE}}(\eta,s)=\sqrt{\frac{2}{\pi}}\eta^{3/2}s^{2}e^{-\frac{1}{2}\eta
s^{2}},
\end{equation}
where $\eta$ is obtained by requiring that $P^{\text{GUE}}$ has a mean spacing
equal 1. Substituting (37) into (32), we obtain the following expression for
the non-extensive (Tsallis) unitary ensemble
\begin{equation}
P^{\text{TsUE}}(m,s)=\sqrt{\frac{2}{\pi}}\frac{\Gamma\left(  \frac{1}%
{2}\left(  3+m\right)  \right)  }{\Gamma\left(  \frac{1}{2}m\right)  }%
\frac{\eta_{2}^{3/2}s^{2}}{\left(  1+\frac{1}{2}\eta_{2}s^{2}\right)
^{(3+m)/2}},
\end{equation}
where
\begin{equation}
\eta_{2}=\frac{8}{\pi}\left[  \frac{\Gamma\left(  \frac{1}{2}\left(
m-1\right)  \right)  }{\Gamma\left(  \frac{1}{2}m\right)  }\right]  ^{2}%
\end{equation}
is obtained by requiring that $\int_{0}^{\infty}sP^{\text{TsUE}}(m,s)ds=1$.
The evolution of the distribution $P^{\text{TsUE}}(s)$\ as $m$ decreases from
$\infty$, where it is given by the Wigner surmise,\ a the limiting value of
$m=2$ is demonstrated in Ref. \cite{abul} for the two dimensional case. Fig. 3
compares the NNSD for the TsUE with the corresponding distributions in the
numerical experiment in Ref. \cite{zyc}. We again see the proposed
non-extensive RMT provides a satisfactory description for the stochastic
transition in terms of a single parameter, particularly in its early stage.

\section{CONCLUSION}

In the present work we have obtained a non-extensive generalization of the
matrix-element theory by extremizing Tsallis's entropy, indexed by $q$,
subjected to two constraints: normalization and finite average norm of the
matrices. We consider the sub-extensive regime of $q>1$, where the transition
from chaos to order described by the Poisson statistics is expected. The
constraint of finite matrix-norm forces an upper limit of the entropic index:,
limiting the attainable range to $1<q<1+2/d$, where $d$ is the dimension of
the matrix-element space. This is essentially the extensive limit since RMT
normally involves large matrices. Nonetheless, the obtained fluctuation
statistics depend mainly on the parameter $m=-d+2/(q-1)$, so that a minor
deviation from extensivity leads to an observable effect. Because of this
limitation, we expect the non-extensive formalism to provide a description for
the early stages of transition from chaos towards regularity. We obtain
distribution functions for the three symmetry universality classes, for which
the probability of larger matrix elements decay algebraically instead of
exponentially. By means of an integral transform, which is based on an
integral representation of the gamma function, we express the characteristics
of the non-extensive theory to those of the standard RMT. We calculate the
level density and the NNSD for systems with mixed regular chaotic dynamics. We
have also derived a generalization for the Wigner surmise that can be compared
to numerical experiments with mixed systems.

Fig. 1. The level density (normalized to 1) for TsOE ploted against energy
eigenvalues expressed in units of $v$ for different values of the parameter
$m$.

\pagebreak%

\begin{figure}
[ptb]
\begin{center}
\includegraphics[
natheight=3.346000in,
natwidth=4.465900in,
height=4.7694in,
width=6.3512in
]%
{IBNZUP00.wmf}%
\end{center}
\end{figure}

Fig. 2. NNSD for TsOE, calculated using Eq. (34) compared with the resuts of
the numerical experiment by K. \.{Z}yczkowski and M Ku\'{s} \cite{zyc}.

\pagebreak%

\begin{figure}
[ptb]
\begin{center}
\includegraphics[
natheight=3.364100in,
natwidth=4.502200in,
height=4.9934in,
width=6.6694in
]%
{IBNZYL01.wmf}%
\end{center}
\end{figure}

Fig. 3. NNSD for TsUE, calculated using Eq. (38) compared with the resuts of
the numerical experiment by K. \.{Z}yczkowski and M Ku\'{s} \cite{zyc}.

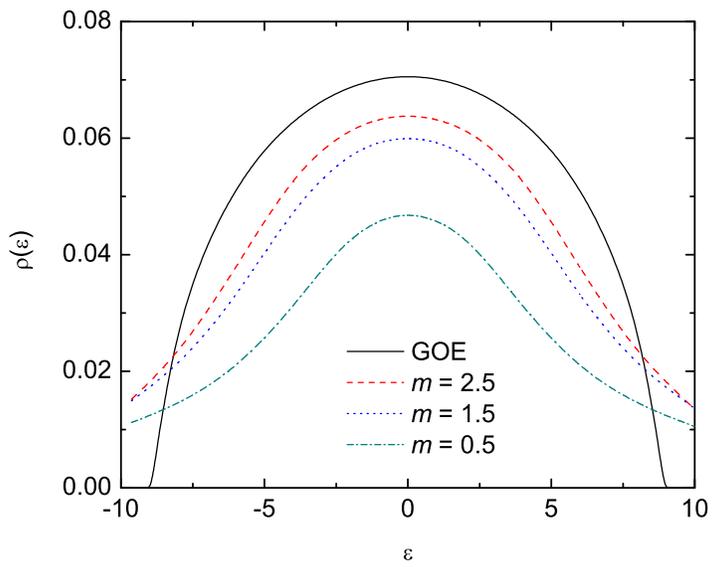
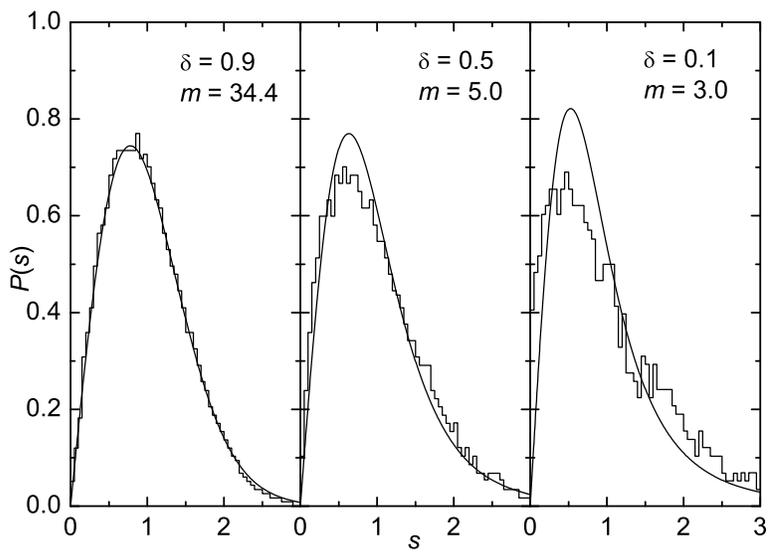
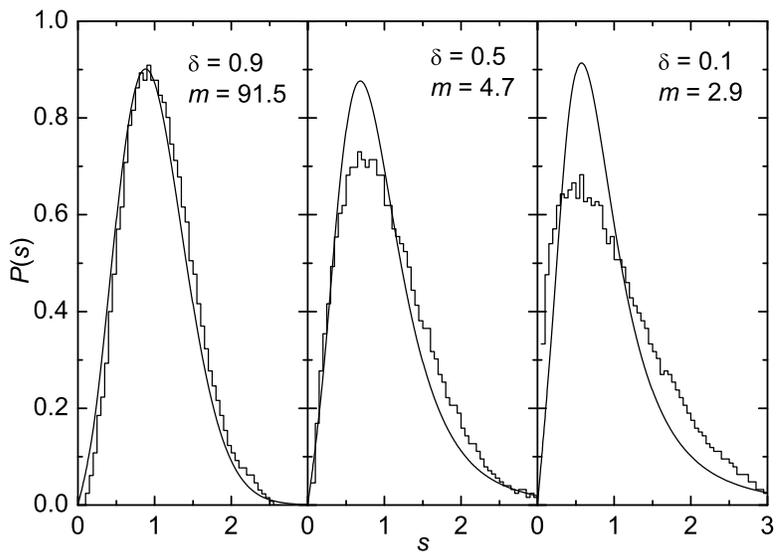
\end{document}